\documentclass[pra,twocolumn,showpacs]{revtex4}

\usepackage[english]{babel}

\usepackage{natbib}
\usepackage{xspace}
\usepackage{dsfont}
\usepackage{times}

\usepackage{graphicx}
\usepackage{amsmath, amsfonts, bbm}

\usepackage{pdfsync}
\usepackage{epsfig, epstopdf,color}

\newcommand{\todo}[1]{}

\renewcommand{\imath}[0]{{\rm i}}

\begin{document}

\title{Temperature dependence of the plasmonic Casimir interaction}

\author{F. Intravaia$^{1,2}$, H. Haakh$^2$ and C. Henkel$^2$}

\address{
$^1$ 
Theoretical Division, MS B213, Los Alamos National Laboratory, Los Alamos, NM 87545, USA\\
$^2$Institut f\"{u}r Physik und Astronomie, Universit\"{a}t Potsdam, Karl-Liebknecht-Stra{\ss}e 24/25, 14476 Potsdam, Germany}

\date{23 April 2010}

\begin{abstract}
We investigate the role of surface plasmons in the electromagnetic
Casimir effect at finite temperature, including situations out of global thermal
equilibrium.
The free energy is calculated analytically and expanded for different regimes of distances and temperatures. Similar to the zero-temperature case, the interaction changes from attraction to repulsion with distance.
Thermal effects are shown to be negligible for small plate separations and at room temperature, but become dominant and repulsive at large values of these parameters. 
In configurations out of global thermal equilibrium, we show that the selective 
excitation of surface plasmons can create a repulsive Casimir force between 
metal plates.
\end{abstract}


\maketitle

\section{Introduction}
The interaction between two parallel plates due to the zero point fluctuations of the electromagnetic field is commonly known as the Casimir effect.
For metallic plates, it is well known that at short distance, the interaction can 
be attributed to surface plasmon modes\cite{Economou69, Raether} that 
hybridize across the vacuum between the interfaces \cite{Van-Kampen68, Gerlach71}.
Surface plasmons have been attracting much interest in the last years in
connection with a broad range of topics such as near-field spectroscopy,
sub-wavelength resolution \cite{BrongersmaBook, MaierBook} or extraordinary optical transmission through subwavelength metallic hole arrays 
\cite{GarciaVidal10,Altewischer02,Fasel05}.
The electromagnetic field associated with these modes is evanescent. 
It therefore came as a surprise that
they also give a large contribution at large distances, and even a repulsive 
one~\cite{Intravaia05a, Intravaia05, Intravaia07}. These papers have been restricted to the Casimir effect at zero temperature. 
The present paper generalizes these results by including a nonzero temperature and situations out of thermal equilibrium. One might expect that the
thermal excitation of surface plasmons is irrelevant, since their typical energies 
are comparable to the plasma frequency of the metal, much larger than 
experimentally relevant thermal energies. As retardation is taken into account,
however, the surface plasmon dispersion relation approaches the light cone
and drops to lower frequencies. These are comparable, for two parallel plates,
to the lowest cavity resonance $\sim c / L $. We find indeed a significant
thermal component to the Casimir interaction between surface plasmon modes
when the distance $L$ exceeds the thermal wavelength 
$\sim \hbar c / k_B T$. By selectively exciting a class of plasmonic modes,
we even get an overall repulsive Casimir force.

The paper is organized as follows.
We first recall the dispersion relations for coupled surface plasmon modes
on two metallic plates~\cite{Economou69} and
obtain a general expression for the corresponding Casimir free energy.
This is expanded asymptotically in different regimes of distance $L$ and temperature $T$  in Sec.\,\ref{asymptotes}.
Sec.\,\ref{entropy} discusses the plasmonic Casimir entropy.
We then compare the results to the full Casimir interaction between metal 
plates (Sec.\,\ref{plasma}),
including all electromagnetic modes, consider situations in which the 
plasmonic modes are not at the same temperature as the rest of the system
(Sec.\ref{s:non-eq}), and conclude with a short summary. 

We adopt throughout this paper the following lossless 
dielectric function
\begin{equation}
\varepsilon(\omega) = 1 - \frac{\omega^{2}_{\rm p}}{\omega^{2}}~,
	\label{eq:epsilon-plasma}
\end{equation}
where $\omega_{\rm p}$ is the plasma frequency.
The Casimir energy and force (both per unit area) are normalized to the
values found for perfectly reflecting mirrors~\cite{Casimir48}
\begin{equation}
\label{eq:Casimir}
E_{\rm C}  =  - \frac{\hbar c}{4\pi \aleph L^3 },\quad 
F_{\rm C}   
=  - \frac{3\hbar c}{4\pi \aleph L^4 }~,
\end{equation}
where 
$\aleph  ={180/\pi ^3 }$.
An intrinsic physical length scale of the system is 
the plasma wavelength 
$\lambda_{\rm p} = 2 \pi c / \omega_{\rm p}$.
It is convenient to use the latter as a length scale, switching to a 
dimensionless plate distance  $\lambda = L / \lambda_{\rm p}$.
We introduce also a reduced temperature 
$ \tau =T/ T_{\rm p}  ={ (2\pi k_B T)}/{\hbar \omega_{\rm p} } $, where
$T_{\rm p}$ is the plasma temperature. 
This choice makes 
the numerical results independent of the specific material, and gives universal
scaling laws. 
Note that $1/\tau$ is proportional to
the ratio between the thermal wavelength (a few microns at room temperature) 
and the plasma wavelength,
and that the product $\lambda \tau = k_B T L / (\hbar c) $
is independent of the plasma wavelength. 
A para\-meter set used frequently in related work is the one for gold: 
$\hbar\omega_{\rm p} 
= 8.96\, \mathrm{eV}$,
$\lambda_{\rm p} = 136\,\mathrm{nm}$ and 
$T_{\rm p} = 1.66\times 10^4 \mathrm{K}$. 
Room temperature then corresponds to $\tau \approx 1.8\times 10^{-2}$.

\section{Plasmonic Casimir free energy}\label{sec:fe}
The Casimir free energy of two metallic plates is obtained by summing the free 
energy (per mode) over the electromagnetic modes vibrating inside the 
cavity~\cite{Van-Kampen68, Casimir48}.
This expression is suitably regularized, namely by subtracting the limit 
of large distances between the plates.

Since the modes of the electromagnetic field are formally equivalent to harmonic oscillators, the free energy of a single mode of frequency $\omega$ in thermodynamical equilibrium at temperature $T$ is
\begin{eqnarray}\label{fe1}
f(\omega) &=&  
\frac{\hbar \omega}{2}
+
k_B T \ln\left[1-e^{-\frac{\hbar \omega}{ k_B T}} \right] 
~.
\end{eqnarray}
%
%
In this paper, we sum Eq.\,(\ref{fe1}) over the dispersion relations for the
surface plasmone modes.
%
\begin{figure}[t]
\centering
   \includegraphics[width=\columnwidth]{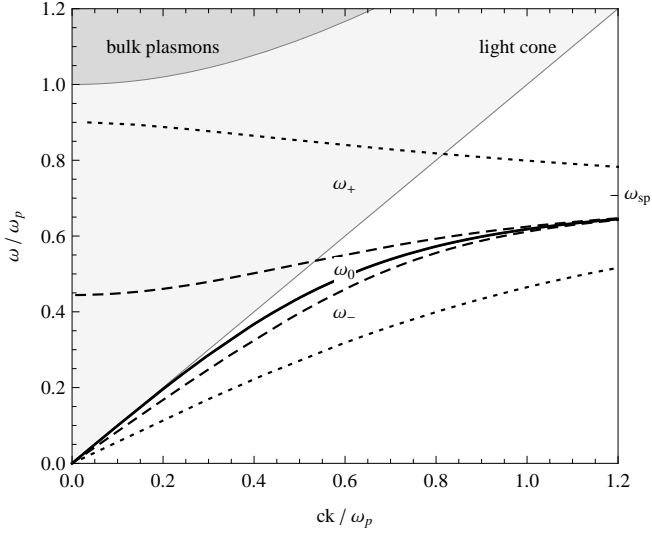}
  \caption{
Dispersion relation of the surface plasmon modes at plate separation 
$L  = c/\omega_{\rm p}$ (dotted curves) and 
$L = 2 c /\omega_{\rm p}$ (dashed curves); $k$ is the wavevector parallel to 
the surface.
The solid black curve $\omega_0( k )$ corresponds to the surface plasmon on 
an isolated plate ($L \to \infty$). The branch $\omega_+( k )$ crosses the 
light cone. The asymptotic value at large $k$ is $\omega_{\rm sp} = \omega_{\rm p} / \sqrt{ 2 }$.
} \label{fig:plasmon_dispersion}
\end{figure}
%
Isolated surfaces (at infinite distance) carry a single surface plasmon mode of frequency $\omega_0( k )$, illustrated in Fig.\,\ref{fig:plasmon_dispersion}.
If the plates are brought together, the electromagnetic fields of the modes
overlap, breaking the degeneracy and splitting the dispersion relation in two branches, whose frequencies we label $\omega_\pm( k )$.
The modes $\omega_-( k )$ and $\omega_0( k )$ are both entirely 
evanescent and lie below the light cone. 
The mode $\omega_+( k )$, however, crosses the light cone and connects
smoothly with the lowest propagating mode (with p-polarization)
within the cavity \cite{Economou69}. 
Adding the free energies of the coupled modes 
and subtracting twice the corresponding values at infinite distance
the integral over the dispersion relations gives
the plasmonic Casimir free energy in the form
(coefficients $c_\pm   = 1$, $c_0  =  - 2$)
\begin{equation}
\label{fe2} 
\mathcal{F}( L, T ) =
\sum_{a =  \pm ,0} c_{a} 
\!\int\limits_{k_a }^\infty \frac{k\, {\rm d}k}{2 \pi}
~  f(\omega_a)
\end{equation}
which is convergent at large $k$ \cite{Intravaia05a}. The thermal
part of the free energy [second term in Eq.\,(\ref{fe1})] naturally cuts off 
modes above $k_B T / \hbar$.
Different choices of the lower limits $k_a$, 
related to the subtraction procedure, are 
possible and have been discussed in Refs.\,\cite{Intravaia05, Intravaia07, Bordag06, Lenac06, Lambrecht08}.
They are connected with 
the way the evanescent and propagating contributions of the mode 
$\omega_{+}( k )$ are split. Here we apply the convention of Refs.\,\cite{Intravaia05,Intravaia07} and set $k_a = 0$ for all modes. We thus
include both propagating and evanescent branches of the `plasmonic mode' 
$\omega_{+}( k )$.

\begin{figure}[b!!]
\centering
   \includegraphics[width=\columnwidth]{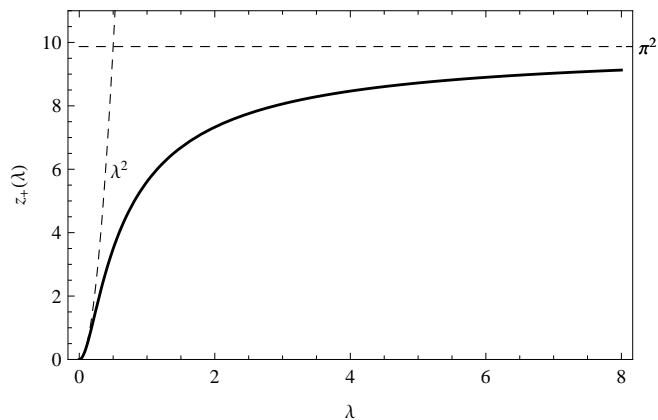}
   \caption{
$z_+(\lambda)$ vs.\ the plate separation $\lambda$ and its asymptotes at small and large distances. The limiting cases are
$z_+ \approx (2\pi\lambda)^2$ and $\approx \pi^2$ for $\lambda \to 0$
and $\to \infty$, respectively.}
\label{fig:zp}
\end{figure}

The calculation of the integral\,\eqref{fe2} is challenging 
because 
the surface plasmon dispersion relations $\omega_{a}( k )$ are solutions of 
a transcendental equation, except in the 
non-retarded limit $k \gg \omega_{\rm p} / c$ where
$\omega_a^2(k) = \omega_{\rm sp}^2( 1 + a \,e^{ - k L }$
with $\omega_{\rm sp} = \omega_{\rm p}/\sqrt{2}$,
$a = 0, \pm$.
Progress can be made with the parametric form described in Refs.\,\cite{Intravaia05a, Intravaia05, Intravaia07}.
Adopting the notation of Ref.\,\cite{Intravaia05},
we get the dispersion relations $\omega_{a}( k )$,
$a \in \{ 0, \pm\}$, from
%
%
%
\begin{equation}\label{fe4}
\omega^2_{a}( z ) = \frac{ c^2 }{ L^2 }
g_{a}^2(z)
, 
\qquad
k^2_{a}( z ) = \frac{ z + g_{a}^2( z ) }{ L^2 }
\end{equation}
with the dimensionless functions
\begin{eqnarray}
g_{a}^2(z)	 &=& \frac{(2\pi \lambda)^2 \sqrt{z}}{\sqrt{z} 
+ \sqrt{z + (2\pi \lambda)^2} \left[\tanh(\sqrt{z}/2)\right]^a}~.
	\label{eq:def-g-a-of-z}
\end{eqnarray}
(The exponents are $\pm 1$ for $a = \pm$.) The parameter $z$ varies
from $- z_{a} \ldots \infty$: one has $z_0=z_- = 0$, while
the number $z_+$,
plotted in Fig.\,\ref{fig:zp} as a function of $\lambda$,
is the solution\cite{Intravaia05} of the
transcendental equation  $\sqrt{z_ +} = 2\pi \lambda \cos( \sqrt{z_ +}/2)$.
The propagating branch of the mode $\omega_+( k )$ corresponds to 
the interval $z \in [-z_+, 0]$; evanescent modes (below the light cone)
have $z > 0$.
%
%
%
Changing the integration variable in Eq.\,(\ref{fe2}) 
from $k$ to $z$, the plasmonic Casimir free energy is given by
\begin{widetext}
\begin{equation}
\label{fe5}
\mathcal{F}   = \frac{\hbar c}{8\pi L^3 } \sum_{a=\pm,0} c_{a}
\left[
 \int\limits_{-z_{a}}^{\infty} 
\left( g_{a}(z)
+2\lambda\tau  \ln\left[1 - e^{ - \frac{g_{a}(z)}{ \lambda\tau}} \right] \right) {\rm d}z+2
 \int\limits_{g_{a}(-z_{a})}^{g_{a}(\infty)}\left(  g^{2}_{a}+2  g_{a}  \lambda\tau \ln \left[ 1 - e^{ -  \frac{g_{a}}{ \lambda\tau}}\right]  \right) {\rm d}g_{a}
\right]
%
\end{equation}
\end{widetext}
For all plasmonic modes, the second integral has a finite upper limit
$g_{a}(\infty) = \sqrt{ 2 }\, \pi \lambda = 
\omega_{\rm sp} L / c$ that coincides with the non-retarded surface plasmon 
frequency; their contributions cancel in the sum over $a$.
The lower boundaries are $g_{0}(-z_{0}) = g_-(-z_-)=0$, and 
$g_{+}(-z_{+})=\sqrt{z_{+}}$.
The first integral in Eq.\,(\ref{fe5}) can only be evaluated approximately
(see Sec.\,\ref{asymptotes}), but a closed form can be given for the second one.

%

In the following, we scale the plasmonic free energy to the 
zero-temperature Casimir value, Eq.\,(\ref{eq:Casimir}),
\begin{equation}
\mathcal{F}(L,T) = E_{\rm C}( L )
	\varphi(\lambda,\tau)
	\label{eq:def-phi}
\end{equation}
and split the correction factor in two terms
\begin{equation}
	\varphi(\lambda,\tau) = 
	\eta( \lambda ) + 
	\vartheta(\lambda,\tau)~,
	\label{eq_eta}
\end{equation}
where the first is the plasmonic Casimir energy at zero 
temperature \cite{Intravaia05, Intravaia07}:
\begin{subequations}
\label{eq:integrals-eta-and-theta}
\begin{eqnarray}
	\eta( \lambda )
&=&  
- \frac{\aleph}{2}\sum_a c_{a} \int_{ - z_{a} }^\infty g_{a}(z) {\rm d}z 
+ \frac{\aleph }{3}z_+^{3/2} ~.
  \label{eq:eta_T1}
\end{eqnarray}
The second term in Eq.\,(\ref{eq_eta}) gives the temperature-dependent 
part for which Eq.(\ref{fe5}) gives
%
\begin{align}
\vartheta(\lambda,\tau)  = 
	 &- \aleph \, \lambda\tau \sum_{a} c_{a} \int_{ - z_{a} }^\infty
	\ln \left[ 1 - e^{ -  \frac{g_{a}(z)}{ \lambda\tau} } \right]{\rm d}z\nonumber\\
	&-  2 \aleph \left( \lambda \tau \right)^3
 	 \mathcal{L} \left( \frac{ \sqrt{z_ +} }{\lambda\tau } \right) 
	 ~,
	 \label{eq:eta_T2}
\end{align}%
\label{thermalfactor}%
\end{subequations}%
where the following combination of polylogarithmic functions appears
\begin{equation}
	{\cal L}( x ) = \zeta( 3 ) 
	- {\rm Li}_3( {e}^{ - x } )
	- x {\rm Li}_2( {e}^{ - x } )
	\label{polifuntion}
\end{equation}
with ${\rm Li}_n(x) = \sum_{k=1}^{\infty} x^k / k^n$.
We note that 
${\cal L}( x ) \sim \frac14 x^2 ( 1 - 2 \log x )$
for small $x$, and ${\cal L}( x ) \to \zeta( 3 )$ exponentially fast for
large $x$.
Eq.\,(\ref{eq:eta_T2}) does not depend only on the product 
$\lambda \tau$ because the material-dependent
parameter $\lambda$ enters via the lower limit $z_+$ and the
functions $g_a( z )$ [Eq.\,(\ref{eq:def-g-a-of-z})].
%

%

\begin{figure}[ht]
\centering
   \includegraphics[width=8cm]{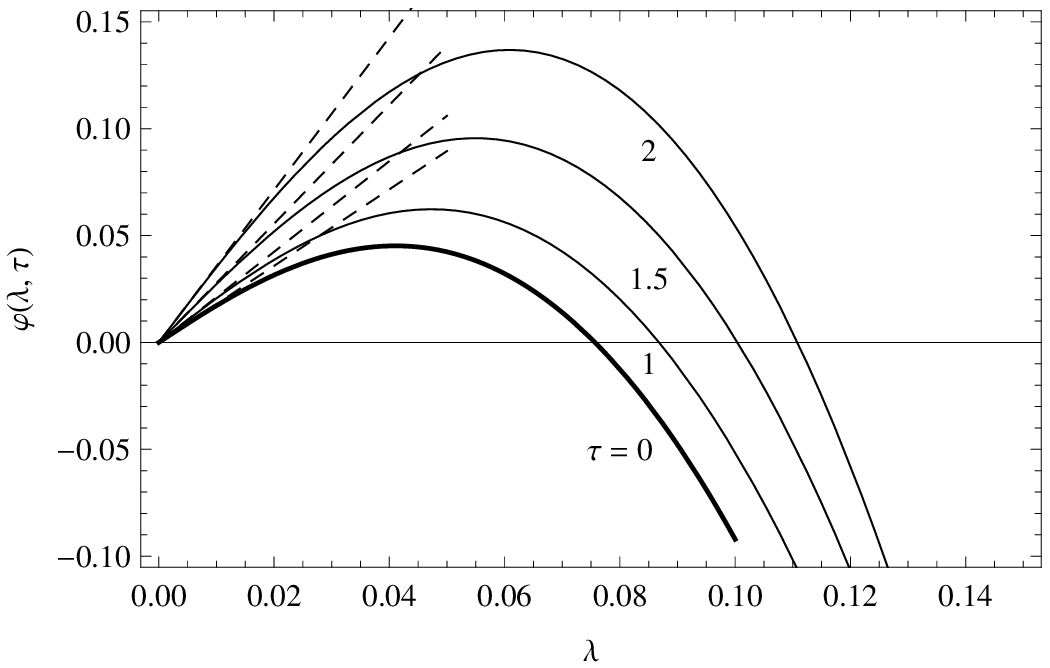}
     \caption{Plasmonic contribution to the Casimir free energy
     vs.\ distance at different temperatures, 
     normalized to the perfect mirror case at $T = 0$
     (energy correction factor $\varphi(\lambda,\tau)$ in Eq.(\ref{eq:def-phi})).
Solid curves: numerical evaluation of Eqs.\,(\ref{eq:eta_T1}, \ref{eq:eta_T2})
for different temperatures;
dashed curves: short-distance limit Eqs.\,(\ref{eq:eta_small_distance}, \ref{TempCorrectionApprox}).
Distance and temperature are scaled to the plasma wavelength
$2\pi c / \omega_{\rm p}$ and temperature $T_{\rm p} 
= \hbar \omega_{\rm p} / k_B$, respectively. 
Negative values correspond to a \emph{repulsive} interaction energy.
}
\label{fig:fe}
\vspace{.5cm}
   \includegraphics[width=8cm]{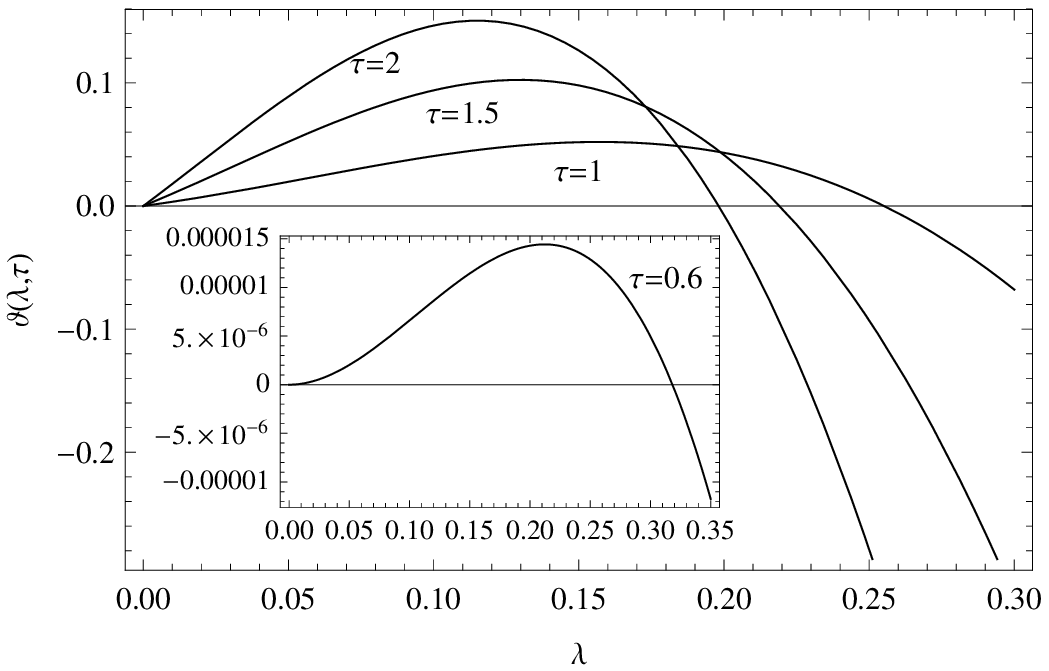}
   \caption{Thermal correction to the plasmonic free energy at short
   distance for different temperatures (energy correction factor
   $\vartheta(\lambda,\tau)$ in Eq.(\ref{eq_eta})).
The behavior qualitatively differs from the complete free 
energy~(Fig.\,\ref{fig:fe}), as a quadratic (rather than linear) distance
dependence emerges at low temperatures, 
cf. Eq.\,\eqref{TempCorrectionApprox}.}
		\label{fig:fe_theta}
\end{figure}
\begin{figure}[ht]
\centering
   \includegraphics[width=\columnwidth]{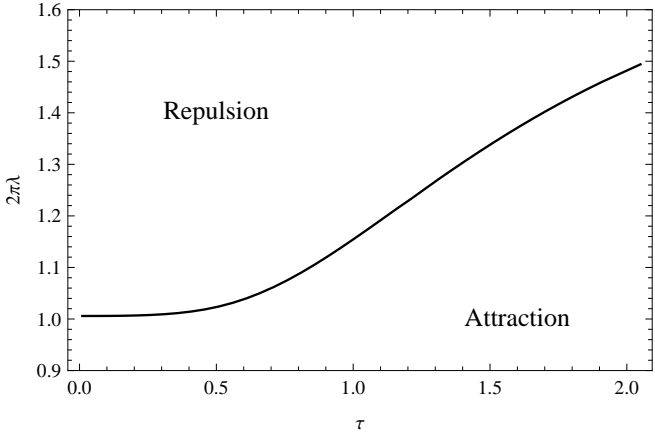}
   \caption{Repulsive and attractive regimes of the plasmonic Casimir pressure in the $(\lambda, \tau)$-plane.
A nonzero temperature slightly increases the distance for which the plasmonic contribution to the Casimir force becomes repulsive.}
		\label{LambdaInv}
\end{figure}

%
The Casimir free energy
is shown in Figs. \ref{fig:fe},  \ref{fig:fe_theta},
\ref{fig:fe_intermediate} and \ref{fig:fe_large} 
for different distance ranges. 
Qualitatively, a nonzero temperature does not modify the behavior of the plasmonic contribution -- we still get a sign change at a distance of order 
$\lambda_{\rm p}/2\pi$, with the interaction becoming repulsive at large
distances. We plot in  Fig.\,\ref{LambdaInv} the inversion distance where 
the Casimir
pressure, $- \partial \mathcal{F} / \partial L$, changes sign:
a weak increase is found as the temperature is raised. Much larger changes
will be found in Sec.\ref{s:non-eq} where configurations out of thermal
equilibrium are discussed.

In the following, we analyze the thermal correction
$\vartheta( \lambda, \tau )$  
in different regimes of distance and temperature.
The zero-temperature Casimir energy 
$\eta( \lambda )$ depends only on one physical scale provided by 
the plasma wavelength and leads to two regimes $\lambda\ll 1$ and 
$\lambda\gg 1$.
For $\vartheta( \lambda, \tau )$ we discuss three regimes.
In all situations of practical interest, it is safe to assume $\tau\ll 1$ for
the scaled temperature and we can distinguish between: 
short distances $\lambda \ll 1$, 
intermediate distances, $1 \ll \lambda \ll 1/\tau$,
and large distances (beyond the thermal wavelength)
$1/\tau \ll \lambda$.

\section{Asymptotic expressions}\label{asymptotes}
%
The calculation of asymptotic expressions requires some care already at zero temperature as was shown in Ref.\,\cite{Intravaia07}.
%
%
When performing approximations on the integrals in Eqs.\,\eqref{eq:eta_T1} and \eqref{eq:eta_T2} one must bear in mind that the functions 
$g_{a}(z)$ cover a wide range of values, from very small to large, depending 
on $z$ and $\lambda$.
Their characteristic scale in the variable $z$ is given by the distance parameter 
$\lambda$.
For $-z_{a}<z \ll (2\pi \lambda)^{2}$ we may use ($a = 0, \pm$)
\begin{equation}
\label{OpSmall}
g_a  \left( z \right) \approx \sqrt{ 2\pi \lambda \sqrt{z} \left[ \coth\left( {\frac{{\sqrt z }}{2}}\right)\right]^{a}}~,
\end{equation}
while for $z\gg(2\pi \lambda)^{2}$  the (non-retarded) approximations
\begin{equation}\label{fe12}
g_a  \left( z \right) \approx
\frac{2\pi \lambda}{ {\sqrt 2 }} \sqrt {1 + a~ e^{ - \sqrt z } }
\end{equation}
hold.
%
It is therefore convenient to split the integration range 
in Eqs. \eqref{eq:eta_T1} and \eqref{eq:eta_T2} as follows
\begin{equation}
\label{intglsplitt}
\int_{-z_a}^{\infty}{\rm d}z=\int_{-z_{a}}^{0}{\rm d}z+\int_{0}^{(2\pi \lambda)^{2}}{\rm d}z+\int_{(2\pi \lambda)^{2}}^{\infty}{\rm d}z~.
\end{equation}
The first integral concerns only the mode $\omega_{+}( k )$ because 
$z_{-}, z_{0} = 0$.
We can use Eq.\,(\ref{OpSmall}) in the first two integrals and 
Eq.\,(\ref{fe12}) in the third. Depending on distance and temperature and on the
desired accuracy, 
these have to be compared with the integrated terms 
in Eq.(\ref{eq:integrals-eta-and-theta})
(proportional to $z_{+}^{3/2}$ or $\mathcal{L}( \sqrt{ z_+ } / \lambda \tau)$).

\subsection{Short distance}\label{short}

At short distance the zero-temperature energy correction was already 
analyzed in Ref.\,\cite{Intravaia07}. It turns out that it is dominated by large 
values of $z$
(third integral of Eq.\,\eqref{intglsplitt}). At the leading order, we get
\begin{equation}\label{eq:eta_small_distance}
\eta(\lambda) \xrightarrow{\lambda \ll 1} 1.790 \lambda~.
\end{equation}
Higher order terms take the form 
$\lambda^{3}(a+b\log \lambda)$ with numerical coefficients
$a$ and $b$ given in the same reference.

Considering the thermal correction for $\lambda\ll1$, the thermal scale becomes important, too. For realistic temperatures, we also have
$\lambda\tau\ll1$, and the main contribution arises from the second and the 
third integral in Eq.\,\eqref{intglsplitt}. Indeed, it can be
shown that the first integral and the polylogarithmic term (involving
$\mathcal{L}$) are beyond the order ${\cal O}( \lambda^2 )$.
In addition, the main contribution to the second integral arises from 
the mode $\omega_{-}( k )$.
This is not surprising since the thermal correction selects frequencies
$\omega_{a}( k ) \lesssim T$
and the mode $\omega_{-}( k )$ is the one that vanishes most quickly as 
$k\to 0$. The corresponding exponent in $k$ determines the power law
in $\tau$, as we discuss at the end of this section.
The opposite case $\lambda \tau \gg1$ is physically irrelevant at short
distances, because one would need $\tau \gg 1$. 
Mathematically,
one finds a divergence from the term proportional to $\mathcal{L}$ 
in Eq.\,(\ref{eq:eta_T2}) that is
exactly balanced by the first integral in Eq.\,\eqref{intglsplitt}. It follows
that the asymptotic form given in Eq.\,(\ref{TempCorrectionApprox}) 
remains valid.

\begin{figure}[ht]
\centering
    \includegraphics[width=\columnwidth]{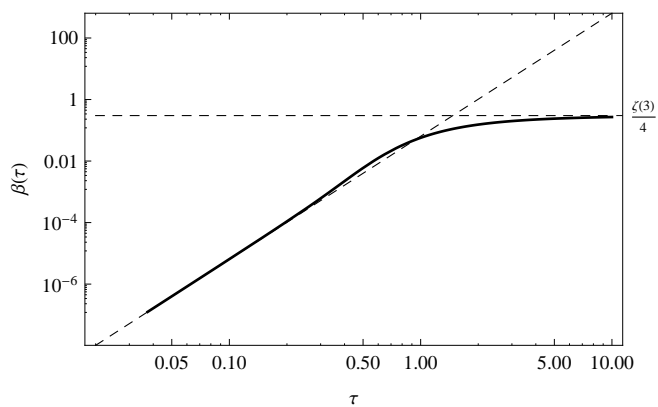}
   \caption{The function $\beta(\tau)$ vs.\ $\tau$ and its asymptotes \eqref{eq:asymptotes_beta} at low and high temperatures.}\label{fig:etahat_small}
\end{figure}

All told, up to the second order in $\lambda$ we find 
\begin{multline}
\vartheta(\lambda,\tau)\xrightarrow{\lambda\ll1} 
\aleph \lambda \tau \left[
2 \frac{\lambda\tau^{2}}{\pi}
\mathcal{L}(2\pi\sqrt{\pi\lambda/ \tau^{2}})
+
\beta(\tau)
\right]
\label{TempCorrectionApprox}
\end{multline}
where the function $\mathcal{L}(x)$ defined in Eq.\,(\ref{polifuntion}) 
appears with a different argument, and where the temperature-dependent
function $\beta(\tau)$ is
 \begin{eqnarray}
\beta(\tau)& =&
\int_0^{\infty}  
\ln \left[
\frac{1- e^{- \pi \sqrt{2}/\tau}}
{1 - e^{ - \pi \sqrt{2 (1 +e^ { -\sqrt z })  }/\tau}}
\right]
 {\rm d}z\nonumber\\
 && +
 \int_0^\infty  
\ln \left[
\frac{1- e^{-\pi \sqrt{2}/\tau}}
{1 - e^{ - \pi  \sqrt{2(1 -e^ { -\sqrt z })  }/\tau}}
\right]
 {\rm d}z ~.
 \label{fe12b} 
 \end{eqnarray}
This is plotted in  Fig.\,\ref{fig:etahat_small} together with its asymptotes 
in the limits of high and low temperatures,
\begin{eqnarray}
\beta(\tau)\xrightarrow{\tau\ll1} 
6\zeta(5) \left(\frac{\tau}{\pi}\right)^{4},  \quad
\beta(\tau)\xrightarrow{\tau\to \infty}
\frac{\zeta(3)}{4}.~
\label{eq:asymptotes_beta}
\end{eqnarray}

We observe the emergence of the characteristic ratio
$\lambda/\tau^{2}$ that determines which of the two
terms in Eq.\,\eqref{TempCorrectionApprox} dominates.
This illustrates that the limits $\lambda\to 0$ and $\tau\to 0$ 
do not commute for the temperature-dependent Casimir energy.
If $\lambda \ll \tau^{2}$ (extremely short distances or high temperatures,
main plot of Fig.\,\ref{fig:fe_theta}), 
the function $\beta(\tau)$ governs the thermal correction 
$\vartheta$ which scales
as $\vartheta \sim \lambda \tau^{5}$ 
if $\lambda\ll\tau^{2}\ll1$. We recover here the same linear distance
dependence as
at zero temperature, Eq.\,\eqref{eq:eta_small_distance}.
The opposite regime $\tau^2 \ll \lambda \ll 1$ emerges at low temperatures,
where the term involving ${\cal L}$ in Eq.\,\eqref{TempCorrectionApprox} 
dominates: we get a behavior
$\vartheta \sim \lambda^2 \tau^3$ (inset of Fig.\,\ref{fig:fe_theta}).
This crossover from a quadratic to a linear scaling with distance can be 
seen in Fig.\,\ref{fig:fe_theta}.

This discussion also illustrates the failure of 
the non-retarded approximation. This leads
to surface plasmon dispersion relations
$\omega^{2}_{\pm}(k) = \omega^{2}_{\rm sp}(1\pm e^{-kL})$ 
and $\omega_{0} = \omega_{\rm sp}$,
and extrapolates a 
free energy $\vartheta \sim \tau^{5}$ down to low temperatures, 
while the correct power is $\tau^{3}$.
This is of course crucial for the low-temperature expansion of 
a thermodynamic quantity like the entropy (see Section \ref{entropy}).

%

%
\begin{figure}[ht!]
\centering
  \includegraphics[width=7.5cm]{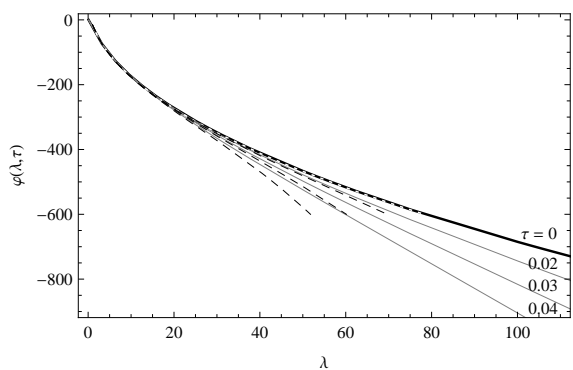}
\caption{Thermal plasmonic Casimir free energy
(reduction factor $\varphi(\lambda, \tau)$ in Eq.(\ref{eq:def-phi}))
 at intermediate distances and different temperatures. Exact numerical calculation (solid curves) and the approximation (\ref{eq:eta_large}) at zero temperature (short-dashed white) and sum of
Eqs.(\ref{eq:eta_large}) and~(\ref{eq:theta_intermediate}) in the 
intermediate regime $\lambda \tau \ll 1$ (long-dashed) respectively .}
\label{fig:fe_intermediate}
\end{figure}

\subsection{Intermediate distance}
	\label{s:intermediate}

%
Parameters for typical experiments are $\tau \approx 10^{-2}$, 
$\lambda \approx 1 \dots 10^{2}$, way beyond short distances.
They lie inside an intermediate regime $1\ll\lambda \ll 1/\tau$, where the 
plate distance is between the plasma and the thermal wavelength. 
Here, the thermal correction to the free energy is still small compared to
zero temperature, as for short distances.

In the scaled Casimir energy $\eta( \lambda, \tau )$,
the main contribution to the integral\,(\ref{eq:eta_T1}) arises for $z\sim 1$ 
and we have to consider the first two integrals in Eq.\,\eqref{intglsplitt},
where, approximately, $z_+ \approx \pi^2$.
%
The energy correction factor at zero temperature then becomes\,\cite{Intravaia07}
%
\begin{eqnarray}\label{eq:eta_large}
 \eta(\lambda) \xrightarrow{\lambda \gg 1}  
 - 74.57 \sqrt{\lambda} + 60. ~,
\end{eqnarray}
%
the offset arising from the second term in Eq.(\ref{eq:eta_T1}).

For the thermal correction, we find the leading order from the polylogarithmic 
$\mathcal{L}$ in Eq.(\ref{eq:eta_T2}).
The $z$-integral gives a contribution which is dominated by 
the interval $z = -z_+ ... (2\pi\lambda)^2$.
Combining the two,
%
\begin{equation} \label{eq:theta_intermediate}
\vartheta( \lambda, \tau )
\approx
- 2\aleph(\lambda\tau)^{3}\zeta(3)
\left( 1 - \frac{1}{\lambda\pi}\right)~.
\end{equation}
%
This gives a small correction that scales as $\tau^3$. 
Both Eqs.\,(\ref{eq:eta_large}) and (\ref{eq:theta_intermediate}) are plotted
in Fig.\,\ref{fig:fe_intermediate}, illustrating the weak impact of temperature.
It is interesting to note that in this range of distances, the thermal plasmonic
contribution is opposite in sign to the free energy of the full electromagnetic
Casimir effect (see Sec.\,\ref{entropy}), and
increases the plasmonic repulsion.
%


\subsection{Large distance}

Let us finally consider the regime $\lambda \gg 1/\tau \gg 1$,
corresponding to a plate separation larger than 
both the plasma and the thermal wavelength. The zero-temperature contribution can still be approximated by 
Eq.\,\eqref{eq:eta_large}, but now the thermal contribution dominates 
the free Casimir energy. The asymptotic behavior of the integrals in 
Eq.\,(\ref{eq:eta_T2}) 
is obtained by expanding the logarithms for small $g_a( z ) / (\lambda \tau)$,
since the functions $g_a( z )$ are bounded:
\begin{equation}
 - \lambda\tau \ln\left[1 - e^{-\frac{g_a(z)}{\lambda\tau}}\right] \approx  
 	- \lambda\tau  \ln\left[\frac{g_a(z)}{\lambda\tau}\right] + 
	\frac{g_a(z)}{2} + \cdots
\end{equation}
Using this expansion under the integral in Eq.\,(\ref{eq:eta_T2}), we note 
that the second term balances exactly with the zero-temperature 
contribution from Eq.\,(\ref{eq:eta_T1}).
As for the first term, we perform the $z$-integration by splitting the
integration range as in Eq.\,\eqref{intglsplitt}.
It is easy to see that in the second interval,
the sum over the mode branches gives zero.
The leading contribution now comes from negative $z$, 
while the third interval in Eq.\,\eqref{intglsplitt} gives an exponentially
small contribution $\sim {e}^{ - 4 \pi \lambda }$.
The polylogarithmic term $\mathcal{L}$ can be expanded for small argument,
which gives eventually
\begin{align}
\label{eq:highT}
 \varphi (\lambda, \tau) 
 \xrightarrow[\lambda \gg 1]{\lambda \tau \gg 1}  
 &- \frac{\aleph \lambda \tau}{2}
 \int_{ - \pi ^2 }^0 \ln \left[ \frac{2\pi }{\lambda \tau^{2}}
\sqrt {z}\coth \left( \frac{\sqrt {z} }{2} \right)  \right]{\rm d}z \nonumber\\
&{} - 2 \aleph 
 \left( \lambda\tau \right)^3\mathcal{L}\left( \frac{\pi }{\tau\lambda }\right)\\
 &\approx
 - \frac{\aleph \pi^{2}\lambda \tau}{2} 
 \left(\ln ( 2\lambda ) -\frac{7\zeta(3)}{\pi^{2}} + \frac{1}{2} \right)
 \nonumber
  \end{align}
where the dependence on $\ln \tau$ cancels to leading order.
%
%
 \begin{figure}[t!]
\centering
   \includegraphics[width=8cm]{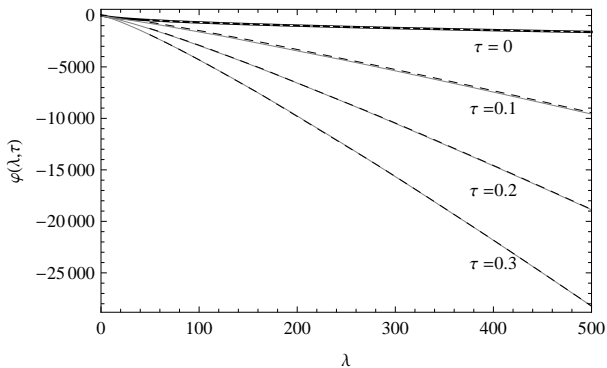}
   \caption{Plasmonic Casimir free energy
   (correction factor $\varphi(\lambda, \tau)$) at large distances and different temperatures. Numerical calculation (solid curves) and large-distance approximations\,(\ref{eq:eta_large}) at zero temperature (white dotted line) 
and~(\ref{eq:highT}) at non-zero temperature (dashed curves).}
\label{fig:fe_large}
\end{figure}
%
The validity range of this asymptotic formula
is illustrated in Fig.\,\ref{fig:fe_large} (dashed lines) where the full
free energy is plotted at large distances.

Summarizing this section, we have generalized
a result known from the zero-temperature case \cite{Intravaia05, Intravaia07} to $T > 0$: 
only the branch 
of the plasmonic mode $\omega_+( k )$ that crosses into the propagating 
sector contributes to 
the (repulsive) plasmonic Casimir interaction at large distances in a significant 
way.




\section{Plasmonic Casimir entropy}
\label{entropy}
\begin{figure}[h!!]
\centering
   \includegraphics[width=8cm]{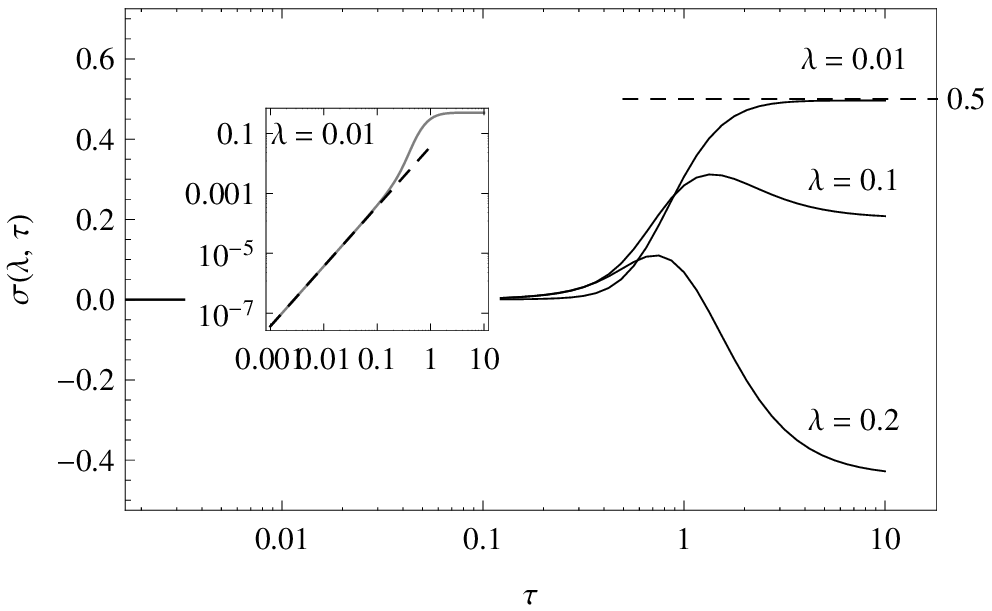}
  \caption{Temperature dependence of the Casimir entropy from plasmonic 
  modes  [correction factor $\sigma( \lambda, \tau )$ relative
   to perfectly conducting mirrors, Eq.\,(\ref{en1})] for short distances and 
high-temperature limit \eqref{shortLhighT} (dashed). 
Inset (double logarithmic scale): low-temperature behavior at short
distance and its 
asymptote (dashed) $\sigma \sim \tau^2$
from Eq.\,\eqref{shortLlowT}.}
  \label{fig:sigma_Small_L}
   \includegraphics[width=8cm]{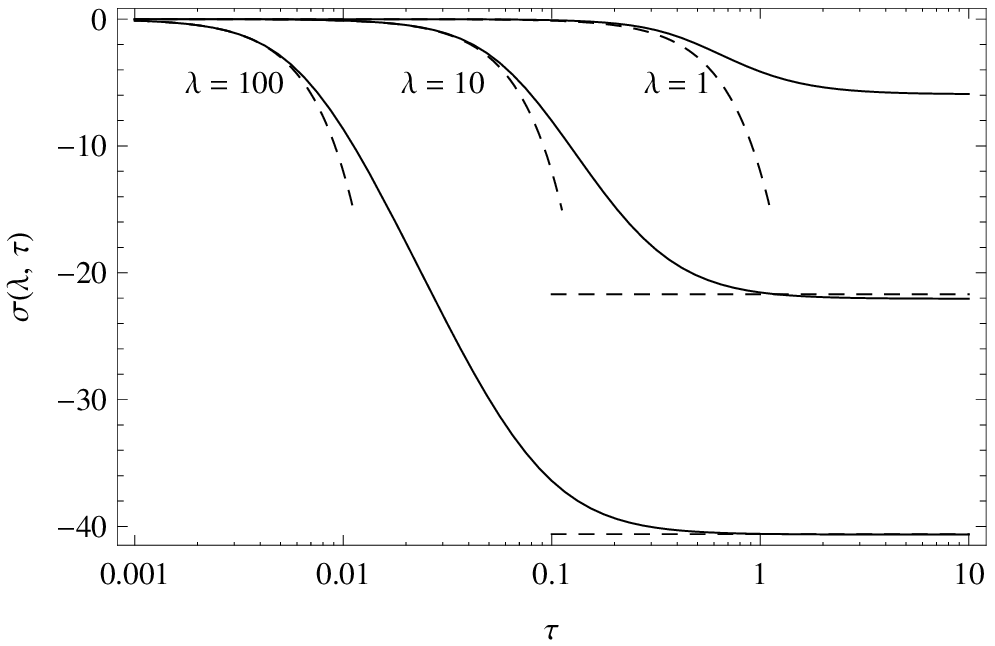}
  \caption{Plasmonic Casimir entropy in the scaled form 
$\sigma(\lambda, \tau)$ [Eq.\,(\ref{en1})],
  vs.\ temperature for intermediate and large distances (solid lines).
Dashed: low- and high-temperature asymptotes 
\eqref{intermediate}, \eqref{eq:entropy_largeL_largeT}.}
\label{fig:sigma_Large_L}
\end{figure}

The plasmonic Casimir entropy can be derived from the plasmonic
Casimir free energy, Eq.\,(\ref{fe5}), by differentiation with
respect to $T$,
\begin{equation}\label{en1}
S(L, T) =  - \frac{{\partial \mathcal{F}}}{{\partial T}}
= S_{\rm C}( L ) \sigma(\lambda, \tau)~.
\end{equation}
A convenient scale is given by the Casimir entropy at high temperatures
between two perfect reflectors (this includes two transverse photon 
polarizations)
\begin{equation}
S_{\rm C}( L ) =\frac{\zeta(3)}{8\pi}\frac{k_{B}}{L^{2}}~.
\end{equation}
The scaled entropy is connected to the dimensionless thermal correction
$\vartheta(\lambda, \tau)$ by a derivative (mind that $E_{\rm C}<0$) 
\begin{equation}
\sigma(\lambda, \tau) = \frac{2}{\zeta(3) \aleph \lambda}
\frac{\partial }{\partial \tau} \vartheta(\lambda,\tau)
~.
	\label{eq:scaled-entropy-and-free-energy}
\end{equation}
%
We recall the result for perfect reflectors where the entropy depends only
on the product $\lambda \tau$\,\cite{Feinberg01}
\begin{equation}
\sigma_{\rm C}(\lambda,\tau) =
\begin{cases}
12 (\lambda \tau)^{2} 
, & \lambda\tau\ll 1\\
1,& \lambda\tau\gg 1.
\end{cases}
	\label{eq:def-perfect-Casimir-entropy}
\end{equation}
The Casimir entropy due to surface plasmons can be represented as the
integral
\begin{widetext}
\begin{equation}
\label{en2}
\sigma(\lambda, \tau) =-\frac{4}{\zeta(3)}
\left(
\sum_a \frac{ c_{a} }{2}
 \int_{ - z_{a} }^\infty
\left( \ln \left[1 - e^{- \frac{g_{a}( z )}{\lambda\tau} } \right] - 
\bar n_a( z ) \frac{g_{a}( z )}{\lambda\tau} \right){\rm d}z 
+3(\lambda\tau)^{2}\mathcal{L}\left(\frac{\sqrt {z_ +}}{\lambda\tau} \right) 
 +
 z_ +
 \ln\left[ 1 - e^{ - \frac{\sqrt{z_+}}{ \lambda\tau}}  \right]
\right)\nonumber~,
\end{equation}
\end{widetext}
where $\bar n_a( z )  = [\exp(g_{a}( z ) / \lambda \tau) -1]^{-1}$ is 
the Bose-Einstein mean photon number.
%
Figs.\,\ref{fig:sigma_Small_L}, \ref{fig:sigma_Large_L} show the 
temperature dependence of $\sigma( \lambda, \tau )$ for several
distances $\lambda$, small and large.
The strong qualitative differences between these cases could be anticipated 
from the free energy of Fig.\,\ref{fig:fe_theta}: at short distances, 
temperature makes increase $\vartheta( \lambda, \tau )$ towards positive 
values, leading to a positive $\sigma$ from 
Eq.\,(\ref{eq:scaled-entropy-and-free-energy}), while the trend is reversed
at larger distances.
%

%
%
Note from Figs.\ref{fig:sigma_Small_L} and \ref{fig:sigma_Large_L}
that the plasmonic Casimir entropy fulfills Nernst's heat theorem 
($\sigma \to 0$ as $T \to 0$) at all distances. Though the same 
is known for the entropy of the complete plasma model \cite{Bezerra02} 
including photonic modes, the result is not trivial because 
surface plasmons are only a subsystem of the two-plate system.

As before one can distinguish three characteristic distance regimes for the plasmonic Casimir entropy.
The expression for $\lambda\ll1$ can be easily obtained from the
approximation\,(\ref{TempCorrectionApprox}) to $\vartheta( \lambda, \tau )$.
At low temperatures (regime $\tau^{2}\ll \lambda\ll1$), we must include
a subleading term in the small-$\lambda$ expansion to get the right
prefactor of the temperature power law. 
This is done by adding to Eq.\,(\ref{TempCorrectionApprox}) the
polylogarithmic term with $\mathcal{L}$ of Eq.\,(\ref{eq:eta_T2})
that becomes
$-2\aleph(\lambda\tau)^{3}\zeta(3)$.
Differentiation leads to
\begin{gather}
 \sigma(\tau, \lambda) \xrightarrow[\tau \ll1]{\lambda\ll1}  \sigma_{\rm C}(\lambda,\tau)\left[\frac{1}{\pi\lambda}+\frac{5}{\pi^{2}} \frac{\zeta(5)}{\zeta(3)}\left(\frac{\tau}{\pi\lambda}\right)^{2} -1\right].
 \label{shortLlowT}
 \end{gather}
The entropy approaches zero quadratically as $\tau\to 0$, as
for perfect reflectors [Eq.\,(\ref{eq:def-perfect-Casimir-entropy})], but the
prefactor is larger by
a factor $1/(\pi \lambda)$. The good agreement with the exact result 
can be seen in the inset of Fig.\,\ref{fig:sigma_Small_L}.

At high temperature, $\sigma(\lambda, \tau)$
becomes a constant that coincides at short distances with the perfect 
reflector limit for one polarization
\begin{equation}
\sigma( \lambda, \tau)\xrightarrow[\tau \gg 1]{\lambda\ll1}  \frac{1}{2}~.
 \label{shortLhighT}
\end{equation}
This can be seen from Eq.\,(\ref{TempCorrectionApprox}) taking into account
the function $\beta(\tau)$.

Intermediate ($1 \ll \lambda \ll 1/\tau$) and 
large ($1, 1/\tau \ll \lambda$) distances can be treated with
Eqs.\,(\ref{eq:theta_intermediate}), (\ref{eq:highT}) that give,
respectively,
\begin{align}
\sigma (\lambda , \tau) &
\xrightarrow[\lambda \gg 1]{\lambda\tau\ll1}
\sigma_{\rm C}(\lambda,\tau)\left( - 1 + \frac{1}{\lambda\pi}\right).
\label{intermediate}
\\
\sigma(\lambda, \tau) &
\xrightarrow[\lambda \gg 1]{\lambda \tau \gg 1}  
- \frac{ \pi^{2}}{\zeta(3)} 
 \left(\ln ( 2\lambda) - \frac{7\zeta(3)}{\pi^{2}}
+ \frac{1}{2} \right)~.
\label{eq:entropy_largeL_largeT}
\end{align}
The validity range of these formulas can be seen from
Fig.\,\ref{fig:sigma_Large_L} (dashed lines). Note that for intermediate
and large distances, the entropy approaches zero from below as $\tau \to 0$.



%
%

\section{Plasmonic vs.\ photonic modes}
\label{plasma}

We now compare the plasmonic Casimir interaction to the full system where
all electromagnetic modes of the cavity are included. The 
knowledge of the surface plasmon contribution provides some physical
interpretation for the complete system. For example,
it is well known \cite{Van-Kampen68, Gerlach71} %
that the full Casimir interaction 
(zero temperature) at small plate separations is well described by taking only
the electrostatic interaction between surface plasmons. 
We now show that this remains valid at nonzero temperature.
The fundamental reason is that at short distances ($\lambda \ll 1$), the 
lowest cavity modes (above the plasmonic $\omega_-+( k )$) have a
characteristic frequency $2\pi c/L$ that already falls in the transparency 
band of the mirrors ($\omega > \omega_{\rm p}$).


%
The full Casimir free energy 
$\mathcal{F}_{\rm Lif} = \varphi_{\rm Lif}(\lambda, \tau) E_{\rm C}$ 
can be obtained from 
the Lifshitz formula \cite{Bezerra02, Lifshitz56}. In our scaled units,
\begin{eqnarray}
\varphi_{\rm Lif}(\lambda,\tau)&=&-2\aleph\lambda\tau \sum_p 
\sideset{}{'}\sum\limits_{n = 0}^{\infty}
\Gamma_p(2 \pi n\lambda\tau)~,\\
\label{TotalFreeEnergy}
\Gamma_p( X ) &=& \int_{X}^{\infty}{\rm d}\kappa\ \kappa
\log[1-r^{2}_{p}(\imath X, \kappa)e^{-2\kappa}]~.
\end{eqnarray}
The index $p \in \{{\rm TE}, {\rm TM}\}$ denotes the polarization. The numbers
$X_n = 2\pi n \lambda\tau$ are scaled Matsubara frequencies.
The Fresnel reflection coefficients in terms of the variables 
$\kappa$ and $X$ are
%
%
%
\begin{equation}
r_{\rm TE}(\imath X, \kappa) = 
\frac{\kappa-\kappa_{m}}{\kappa+\kappa_{m}},
\quad 
r_{\rm TM}(\imath X, \kappa) = 
\frac{\tilde \varepsilon({\rm i}X)\kappa-\kappa_{m}}{\tilde \varepsilon({\rm i}X)\kappa+\kappa_{m}}
\end{equation}
with 
$\kappa_{m}= \sqrt{\kappa^{2} + (2\pi\lambda)^2 }$ 
and the dielectric function of the plasma 
model [cf.\ Eq.\,(\ref{eq:epsilon-plasma})]
$\tilde\varepsilon({\rm i} X ) = 1 + (2\pi\lambda / X)^2$.

 \begin{figure}[h]
\centering
   \includegraphics[width=8cm]{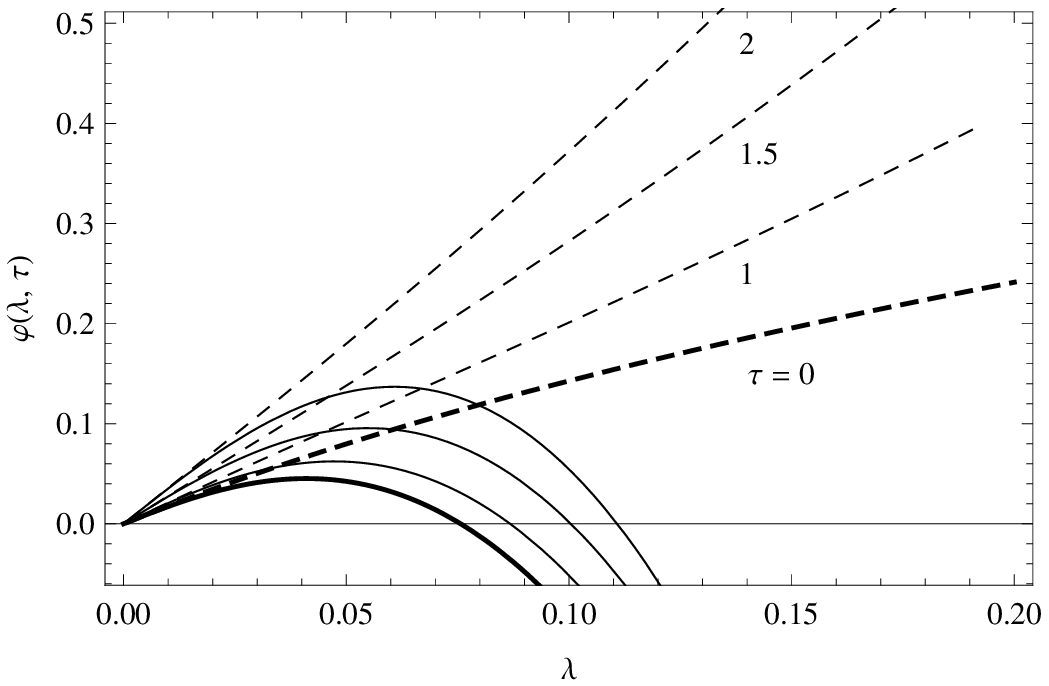}
\caption{Casimir free interaction energy including all modes (Lifshitz
theory with the plasma model, dashed lines), compared to its 
plasmonic counterpart (solid). All energies are 
expressed relative via the correction factor $\varphi( \lambda, \tau )$.
}
\label{fig:plasmaFE}
   \includegraphics[width=8cm]{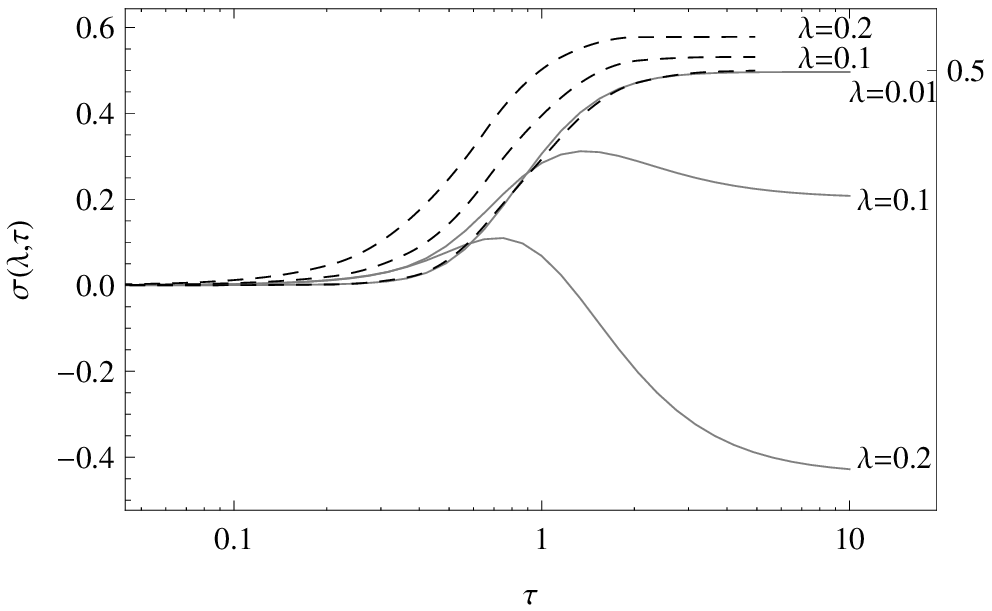}
\caption{Casimir entropy vs.\ temperature for all modes (plasma model,
dashed), compared to the contribution of plasmonic modes only 
(solid). Entropies expressed via the correction factor $\sigma( \lambda, \tau )$.
}\label{en}
  \label{fig:sigma_plasma_Small_L}
\end{figure}

Fig.\,\ref{fig:plasmaFE} shows the scaled free energies for both the 
full Casimir interaction (dashed) and the plasmonic contribution alone (solid).
A good 
agreement is visible at short distances even at nonzero temperature.
In this regime, we can, therefore, obtain detailed information on the 
thermodynamics of the Casimir effect by just considering the plasmonic contribution which can be worked out in analytic form quite easily.
For example, we can immediately conclude that the (full) Casimir entropy at 
$\lambda \ll 1$ is given by Eq.\,\eqref{shortLlowT} at low temperatures
($\tau \ll 1$) and by Eq.\,(\ref{shortLhighT}) for $\tau \gg 1$.

%
%
With respect to the Casimir entropy, 
Figs.\,\ref{fig:plasmaFE} and \ref{fig:sigma_plasma_Small_L} illustrate that the 
Lifshitz expression deviates significantly from the plasmonic 
contribution when 
$\lambda \gtrsim 0.1$. It is obvious that propagating (photonic) modes 
then become relevant.
%
In Appendix\,\ref{propPlasma} we calculate their contribution at low
temperatures for the plasma model:
\begin{equation}
\sigma_{\rm ph}(\lambda,\tau) 
\xrightarrow[\tau \ll 1]{\lambda\tau\ll1} 
\sigma_{\rm C}(\lambda,\tau)
\left[2-\frac{8\pi^{2}\tau}{45\zeta(3)}\frac{2+\pi\lambda}{3}\right]
\label{shortlowProp}
\end{equation}
where the first term is twice the value obtained for perfect 
mirrors.
%
%
%
This is precisely compensated by the plasmonic contribution. Indeed,
for intermediate distances $1 \ll \lambda \ll 1/\tau$,
both Eqs.\,(\ref{intermediate}) and (\ref{shortlowProp}) are valid, and
their sum reproduces the entropy of the full Casimir effect calculated in 
Refs.\,\cite{Bezerra02,Bezerra02a,Bezerra04,Brevik06}.
Evaluating Eq.\,(19) of Ref.\,\cite{Bezerra04} in the regime of intermediate distances, we have
\begin{equation}\label{eq:bezerra}
\sigma_{\rm Lif}(\lambda,\tau)
\xrightarrow[\lambda \gg 1]{\lambda\tau\ll1}
\sigma_{\rm C}(\lambda,\tau)\left[1+\frac{1}{\pi\lambda}- \frac{8\pi^{2}\tau}{45\zeta(3)}\frac{\pi\lambda + 2}{3}\right]
\end{equation}
%


\section{Beyond thermal equilibrium}
\label{s:non-eq}

Until now, we have assumed both metallic slabs to be at the same temperature $T$.
The previous results enable us to deal 
in a simple way with a more general situation, too, where each of the (otherwise 
identical) slabs is described by a local temperature $T_{1}$ and $T_{2}$.
The general theory in this case was investigated in Refs.\,\cite{Dorofeyev98, 
Antezza08}: the non-equilibrium Casimir 
interaction for a symmetric cavity is obtained by simply averaging over the 
equilibrium free energies of the two mirrors
\begin{equation}
\mathcal{F}^{\rm neq}(L, T_{2},T_{1})=\frac{1}{2}\left[\mathcal{F}^{\rm eq}(L, T_{2})+\mathcal{F}^{\rm eq}(L, T_{1})\right]~.
	\label{non-eq}
\end{equation}
%
(In Ref.\,\cite{Antezza08}, this result was derived for the pressure, but 
the same reasoning can be applied for the free energy.)
Combining this formalism with the results from the present paper, it is straightforward to 
calculate the plasmonic contribution to the non-equilibrium Casimir free energy.
Using the split~(\ref{eq_eta}) of the plasmonic free energy into a 
zero-temperature and a thermal part, Eq.\,(\ref{non-eq})
gives for two slabs at different temperatures
%
\begin{equation}
 \varphi^{\rm neq}(\lambda, \tau_{2}, \tau_{1}) =\eta(\lambda)  +   
 \frac{1}{2} \left[\vartheta(\lambda,\tau_{2})+ \vartheta(\lambda,\tau_{1}) \right]
.
\end{equation}
From the results given above, we conclude that qualitatively, 
the behavior of $\varphi^{\rm neq}(\lambda, \tau_{2}, \tau_{1})$ 
is similar to the equilibrium configuration, including a change in sign of the force 
with the distance. This is also confirmed by the asymptotic expressions for 
long/short distance and low/high temperature that can be easily extracted 
from the above results. 
The total Casimir force between identical plates, however, is always attractive,
as is known from Refs.\,\cite{Dorofeyev98,Antezza08} for all temperatures.

\begin{figure}[h]
\centering
   \includegraphics[width=\columnwidth]{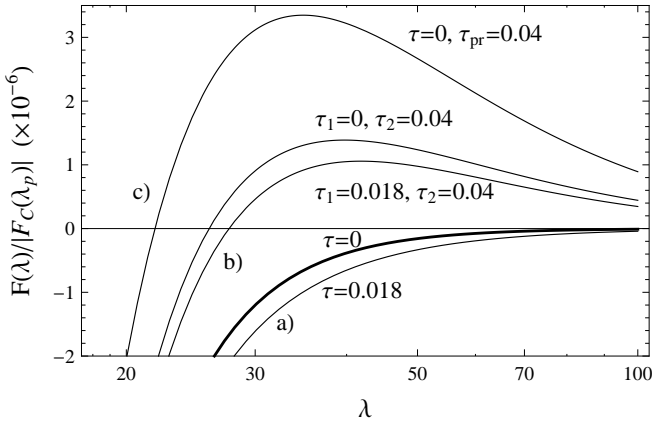}
  \caption{
  Total Casimir force (per unit area) in thermal equilibrium (thick line)
  and in different
  non-equilibrium scenarios. The force is normalized to 
  $10^{-6} |F_{\rm C} (\lambda_{\rm p})|$, approximately
  $3.65\,\mu\mathrm{Pa}$ for gold ($\lambda_{\rm p}=136 \mathrm{nm}$).
 In these units, $T = 300\,{\rm K}$ ($665\,\rm K$) corresponds to 
 $\tau \approx 0.018$ ($0.04$), respectively.\\
a) Total equilibrium at temperature $\tau$: attractive pressure at all distances.
b) Surface plasmon modes of one plate out of equilibrium at 
temperature $\tau_2$,
all other modes at temperature $\tau_1$.
c) All modes at $\tau$, except for the propagating branch of the plasmonic
 mode ($\omega_{+}( k )$) which is at temperature $\tau_{\rm pr}$.
The sign change to repulsion (positive pressure) would occur for gold
at distances between $\approx 2.7\,\mu$m and 
$3.7\,\mathrm{\mu m}$.}
\label{fig:NEQ-zero-T}
\end{figure}

Let us now consider a slightly different non-equilibrium scenario where temperature
is still raised locally (in one plate), but only for a subclass of modes.
If it were possible to increase the mean excitation of the plasmonic modes on one plate,
above the equilibrium level of the propagating (photonic) modes, 
the total Casimir free energy would read
\begin{equation}
\varphi^{\rm neq}_{\rm Lif}(\lambda, \tau_{2}, \tau_{1}) =
\varphi_{\rm Lif}(\lambda, \tau_{1}) +\frac{1}{2}\left[\vartheta(\lambda,\tau_{2})- \vartheta(\lambda,\tau_{1}) \right]~,
	\label{eq:neq-all-plasmons-at-T2}
\end{equation} 
where the first term is the total Casimir free energy at equilibrium.
The set of curves b) in Fig.\,\ref{fig:NEQ-zero-T} illustrates that this scenario can create a regime
where the \emph{total} Casimir force becomes repulsive, and this over a fairly
large range of distances. We plot the Casimir pressure (non-equilibrium
force per unit area) when the photonic modes are either at the scaled
temperature $\tau_1$
(zero or room temperature),
and the plasmonic modes on plate 2 at $\tau_2 > \tau_1$. It appears that 
this setting breaks the delicate
balance between photonic and plasmonic modes we
found in Sec.\,\ref{plasma}. A similar interpretation has been put forward
in Ref.\,\cite{Pitaevskii05a}
for the change in distance dependence
of the atom-surface interaction out of equilibrium.
The two values for $\tau_1$ give 
close results because in the intermediate distance range, the effect of the 
temperature is still moderate for the equilibrium case. 
As could be expected, the inversion distance increases and the maximal repulsion
becomes weaker as $\tau_{1}$ increases towards $\tau_2$.
%

We have also included in Fig.\,\ref{fig:NEQ-zero-T} a scenario where 
only
the propagating part of the plasmonic mode $\omega_+( k )$ is
populated at a temperature different from the rest of the system.
It contributes a free energy
\begin{align}
\vartheta^{\rm pr}_{\rm +}(\lambda,\tau)  = 
	 &- \aleph(\lambda\tau)  \int_{ - z_{+} }^0
	\ln \left[ 1 - e^{ -  \frac{g_{a}}{ \lambda\tau} } \right]{\rm d}z
		\label{eq:propagating-plasmon-free-energy}
\\
	&-  2 \aleph \left( \lambda \tau \right)^3 
	\left[
 	 \mathcal{L} \left( \frac{ g_+( - z_ + ) }{\lambda\tau } \right)
	 -\mathcal{L} \left( \frac{ g_+( 0 ) }{\lambda\tau } \right)
	\right]
	,
\nonumber
\end{align}
where $g_+( - z_ + ) = \sqrt{ z_+ }$ as noted before, and
\begin{equation}
g_+(0) = 2\pi \lambda\sqrt{\frac{1}{1+\pi \lambda}}
\end{equation}
gives the (dimensionless) wavevector for which the dispersion relation 
$\omega_{+}( k )$ crosses the light cone.
This leads to a non-equilibrium free energy
%
\begin{equation}
\tilde{\varphi}^{\rm neq}_{\rm Lif}(\lambda, \tau_{\rm pr}, \tau) 
= \varphi_{\rm Lif}(\lambda, \tau) 
+ \vartheta^{\rm pr}_{\rm +}(\lambda,\tau_{\rm pr}) 
- \vartheta^{\rm pr}_{\rm +}(\lambda,\tau)~,
	\label{eq:scenario-hot-propagating-branch}
\end{equation} 
where $\tau_{\rm pr}$ is the temperature of the propagating plasmons
and $\tau$ is the temperature of all other modes. The corresponding 
pressure [curve c) in Fig.\,\ref{fig:NEQ-zero-T}] 
increases with respect to the previous non-equilibrium scenario 
by approximately a factor of $2$, 
and repulsion sets in at a somewhat
shorter distance. This is because 
(i) the -- otherwise attractive -- mode $\omega_-(k)$ is less excited and
(ii) the propagating branch of $\omega_+(k)$ dominates the
interaction at these distances and excites the electron plasma on both 
plates rather than a single one.


The selective excitation of surface plasmon modes is a well-studied problem
(Ref.\,\cite{Renger09} and references therein). 
Most of the setups have to 
cope with the fact that the corresponding electromagnetic field is evanescent 
and, therefore, cannot be excited directly by laser photons incident from 
free space. Corrugated surfaces are of some help here 
\cite{Raether}, and indeed they can convert thermally excited plasmons into far-field
radiation \cite{Greffet02a}.
A recently developed four-wave mixing 
scheme permits to excite surface plasmons even on flat 
surfaces \cite{Renger09}.
The non-equilibrium situation involving only propagating modes 
[Eq.\,(\ref{eq:scenario-hot-propagating-branch})]
may be simpler to realize experimentally since these modes couple 
to free-space light fields and can in principle be excited by shining a
laser \cite{Povinelli} from the side onto the gap between the mirrors. (See 
Ref.\,\cite{Ovchinnikov00} for a related discussion.)



\section{Discussion and conclusion}

We have calculated the contribution to the thermal Casimir effect due to 
surface plasmons, which are hybrid field-matter eigenmodes of metallic surfaces.
The expression we found for the free energy of interaction is valid at any
distance and temperature, and we have derived its asymptotics at small,
intermediate, and large distances.
Thermal effects become significant when the distance is larger than the 
thermal wavelength $\lambda_T$, similar to perfectly conducting plates, 
and below $\lambda_T$ for non-equilibrium configurations.
The other 
length scale of the system (plasma wavelength) determines the
detailed behavior of the free energy.

We have found that at short distances and temperatures the thermal correction 
is small, and that the plasmonic Casimir interaction changes sign with distance, 
leading to a repulsive regime, as has been known from zero 
temperature \cite{Intravaia05, Intravaia07}.
This goes hand in hand with a change of sign of the plasmonic Casimir 
entropy.
In the short-distance regime, we found that the complete Casimir interaction 
between metallic plates (described by the plasma dielectric function) is 
completely dominated by the surface plasmon contribution.
The asymptotic scaling laws explain why $T = 0$ is a good approximation in 
most experimentally relevant situations (intermediate distance regime, low 
temperature). In this regime the known result for the complete plasma model is 
recovered in a simple way by adding propagative photonic modes.

Things are different at high temperatures and large distances. Here it was shown that the plasmonic Casimir interaction is determined by a branch of the 
surface plasmon dispersion relation corresponding to propagating modes, 
resulting in a large repulsive contribution that is enhanced by the temperature.
This effect is probably one of the best illustrations of Casimir repulsion that arises
from the radiation pressure of a standing wave mode. The pressure is repulsive,
because the travelling photons are bouncing off the cavity walls, while 
the reference mode, a single-interface plasmon, has an 
evanescent field with zero radiation pressure.

The balance between plasmonic and photonic modes was emphasized by 
considering two configurations out of global thermal equilibrium where
plasmonic modes are selectively excited to a higher temperature.
These configurations show a crossover to a total Casimir force 
that becomes
repulsive at plate distances $L\approx 20\,\lambda_{\rm p} 
\dots 25\,\lambda_{\rm p}$ (a few microns for gold). This can be
understood qualitatively in terms of radiation pressure due to the 
propagating branch of the plasmonic mode. We emphasize
that this happens at distances shorter than the thermal wavelength where 
the Casimir pressure is stronger.

In conclusion, it seems in principle possible to tune the sign of the Casimir force by 
the selective excitation of the surface plasmons. Still, future research must address 
experimentally relevant questions for such a scheme, e.g.\
how to avoid exciting photonic modes just above the plasmonic one and
how to populate plasmonic modes over a wide angular range.

\paragraph*{Acknowledgments.}
We would like to thank H. T. Dinani and S. Slama for discussions and help
with some calculations.
We benefited from exchanging ideas within the Research Network
``Casimir'' of the European Science Foundation (ESF).  
FI acknowledges partial financial support by the Humboldt foundation and LANL.
HH and CH acknowledge funding by the German-Israeli Foundation for Development and Research (GIF).
\appendix
\section{Full Casimir entropy at low temperatures}
\label{propPlasma}


The dimensionless correction factor for the
Casimir entropy of the plasma model can be written 
as the following integral over (scaled) real frequencies
\begin{equation}
\sigma_{\rm Lif}( \lambda, \tau )
= -\frac{4}{\pi\zeta(3)}\int_{0}^{\infty} \frac{ x\, {\rm d}x }{\sinh^{2}x}
{\rm Im}\sum_{p}M_{p}(2 x \lambda\tau)
\label{entropyPlasma}
\end{equation}
where $p = {\rm TE, \, TM}$ indicates again the polarization and
\begin{align}
M_{p}(\Omega)& = 
\int_{0}^{\infty}\!{\rm d}\kappa\ \kappa
\ln\left[1-r^{2}_{p}(\Omega,\kappa)e^{-2\kappa}\right]
\nonumber\\
&
+\int_{0}^{\Omega}\!{\rm d}y\ y 
\ln\left[1-r^{2}_{p}(\Omega, -{\rm i} y)e^{2\imath y}\right]
	\label{eq:def-M-full-Casimir}
\end{align}
The first (second) integral in Eq.\,(\ref{eq:def-M-full-Casimir}) corresponds
to the evanescent wave (propagating wave) sector, respectively.
For $p={\rm TE}$, 
the argument of the logarithm is always positive in the first integral,
hence its imaginary part vanishes. 
%
This does not happen for $p={\rm TM}$ 
where the first integral gives the contribution of surface plasmons (evanescent
branch) which has been evaluated in this paper.
As mentioned in Sec.\,\ref{plasma}, we are interested here in
the propagating contribution only.

The function $x/\sinh^{2}x$ significantly differs from zero only for 
$x \lesssim 1$. In the limit $\lambda\tau\ll1$, we can therefore expand 
the integrands in $M_p(\Omega)$ for small $y$ and $\Omega$ since
$y \le \Omega \ll 1$. This yields
%
\begin{gather}
{\rm Im}\ M_{\rm TE}^{\rm ph}(\Omega) 
\approx 
-\frac{\pi}{4}\Omega^{2}+\frac{\Omega^{3}}{3\pi\lambda}
\left(1+\pi\lambda\right)
\\
{\rm Im}\ M_{\rm TM}^{\rm ph}(\Omega) \approx 
-\frac{\pi}{4}\Omega^{2}+\frac{\Omega^{3}}{3\pi\lambda}
\left(3+\pi\lambda\right)
\end{gather}
Performing the $x$-integral in Eq.\,\eqref{entropyPlasma},
%
\begin{equation}
\sigma_{\rm ph}(\lambda,\tau) \xrightarrow [\tau\ll1]{\lambda\tau \ll1}\sigma_{\rm C}(\lambda,\tau)
\left[1-\frac{8\pi^{2}\tau}{45\zeta(3)}\frac{2+\pi\lambda}{3}\right]
	\label{eq:result-propagating-modes}
\end{equation}

Note, however, that this result contains the propagating branch of the 
plasmonic mode $\omega_{+}( k )$ whose free energy is given by
Eq.\,(\ref{eq:propagating-plasmon-free-energy}). 
Reviewing the analysis from Secs. \ref{short}, \ref{s:intermediate}, 
it is easy to see that in the limit considered here,
the polylogarithmic term dominates 
in Eq.\,(\ref{eq:propagating-plasmon-free-energy}) and becomes
$\mathcal{L}( \sqrt{z_+} / \lambda\tau) \approx \zeta(3)$.
Subtracting this contribution from\,(\ref{eq:result-propagating-modes}),
we find the entropy of the propagating photonic  
modes given in Eq.\,\eqref{shortlowProp}.


\end{document}